\documentclass[twocolumn,showpacs,floats,floatfix,aps,pra]{revtex4}
\usepackage{amsfonts,amssymb,amsmath}
\usepackage{color,calc}
\usepackage[dvips]{graphicx}
\usepackage{bm,bbm}

\def\be{\begin{equation}}
\def\ee{\end{equation}}
\def\bea{\begin{eqnarray}}
\def\eea{\end{eqnarray}}
\def\bse{\begin{subequations}}
\def\ese{\end{subequations}}

\def\i{\text{i}}
\def\e{\text{e}}

\def\1{\mathbf{1}}

\begin{document}

\author{G. T. Genov}
\affiliation{Department of Physics, Sofia University, James Bourchier 5 blvd, 1164 Sofia, Bulgaria}
\author{A. A. Rangelov}
\affiliation{Department of Physics, Sofia University, James Bourchier 5 blvd, 1164 Sofia, Bulgaria}
\author{N. V. Vitanov}
\affiliation{Department of Physics, Sofia University, James Bourchier 5 blvd, 1164 Sofia, Bulgaria}
\title{Propagation of light polarization in a birefringent medium: Exact analytic models}
\date{\today}

\begin{abstract}
Driving the analogy between the coherent excitation of a two-state quantum system and the torque equation of motion, we present exact analytic solutions to different models for manipulation of polarization in birefringent medium. These models include the one-dimensional model, the Landau-Zener model, and the Demkov-Kunike model. We also give an example for robust, broadband manipulation of polarization by suitably tailoring the birefringence vector.
\end{abstract}

\maketitle


\renewcommand{\textfraction}{0.1}
\renewcommand{\topfraction}{1.0}
\renewcommand{\bottomfraction}{1.0}
\renewcommand{\floatpagefraction}{0.7}

\section{Introduction}

Various problems in many branches of physics are described by torque equations. These include the textbook examples of the classical motion of a charged particle in  a magnetic field \cite{Jackson}, the motion of a point mass under the Coriolis force as exemplified by the behavior of a gyroscope acted on by gravity \cite{Sommerfeld}, and the dynamics of a spin in a magnetic field \cite{Slichter,AE}. More recent applications include  the dynamics of optical waveguides \cite{Longhi}, polychromatic beam splitters \cite{Dreisow}, sum frequency conversion techniques in nonlinear optics \cite{Suchowski08,Suchowski09}, transformation of light polarization in a birefringent medium \cite{Kubo80,Kubo81,Kubo83,Sala,Gregori,Tratnik,RangGauVit}, and other nonlinear processes of three wave mixing \cite{Boyd}.

In this paper we explore the analogy between the torque equation of motion and the coherent dynamics of a two-state quantum system (spin $\frac12$ or a two-state atom) to describe analytically the evolution of light polarization in a birefringent medium by using the approach of Bloch \cite{Bloch} and Feynman \emph{et al.} \cite{Feynman}. We use this approach to find the propagator for the Stokes polarization vector, given the propagator for the respective two-state system. This result allows us to find the conditions for some useful manipulations of the polarization. Then, we derive three exact analytic solutions for (i) the one-dimensional (resonance) model, (ii) the Landau-Zener model \cite{Landau,Zener}, and (iii) the Demkov-Kunike model \cite{RZ,DK,AE,VitanovARZ}.

\begin{figure}[h]
 \begin{center}
 \includegraphics[width=.4 \textwidth]{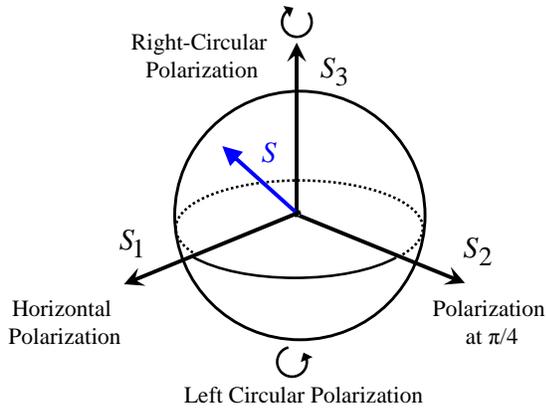}
 \end{center}
\caption{Depiction of the Stokes vector on the Poincare sphere.}
 \label{Poincare}
\end{figure}

This paper is organized as follows. In section \ref{Sec2} we show the analogy between the torque equation for the Stokes polarization vector and coherent excitation of a two-state quantum system. In section \ref{Sec3} we present several exact solutions for polarization evolution, in which the birefringence vector elements correspond to the one-dimensional (resonance) model, the Landau-Zener model, and the Demkov-Kunike model.
Within the latter two models, we discuss a potentially important example for broadband manipulation of polarization.

\section{Evolution of polarization} \label{Sec2}

\subsection{Stokes polarization vector}

Consider a plane electromagnetic wave traveling in the $z$ direction through an anisotropic dielectric medium without polarization dependent loss (PDL). The light polarization is described by the Stokes polarization vector $\mathbf{S}(z)=[S_{1}(z),S_2(z),S_{3}(z)]^T$, which is conveniently depicted on the Poincar\'{e} sphere shown in Fig. \ref{Poincare}. The three basic polarizations are described as follows:
\begin{subequations}
\begin{align}
 \text{right circular:}&\quad \mathbf{S}(z)=[0,0,1]^T; \label{right}\\
 \text{left circular:}&\quad \mathbf{S}(z)=[0,0,-1]^T; \label{left}\\
 \text{linear:}&\quad \mathbf{S}(z)=[\cos 2\phi,\sin 2\phi,0]^T\label{linear};
\end{align}
\end{subequations}
the elliptical polarization is described by the points between the poles and the equatorial plane.

The polarization evolution obeys the following torque equation for the Stokes polarization vector \cite{Kubo80,Kubo81,Kubo83,RangGauVit}:
\begin{equation}
\frac{\,\text{d}}{\,\text{d}z}\mathbf{S}(z)=\mathbf{\Omega }(z)\times \mathbf{S}(z),  \label{Stokes equation}
\end{equation}
where $\mathbf{\Omega }(z)=[\Omega _{1}(z),\Omega _2(z),\Omega _{3}(z)]^T$ is the birefringence vector of the medium; the direction of $\mathbf{\Omega }(z)$ is given by the \emph{slow} eigenpolarization and its length $|\mathbf{\Omega }(z)|$ corresponds to the rotary power.
The torque equation (\ref{Stokes equation}) comprises three coupled linear differential equations for the (real) components of the Stokes vector $S_{k}(z)$ $(k=1,2,3)$. To this end, we use the formal analogy of the Stokes polarization equation (\ref{Stokes equation}) with the Bloch equation in quantum mechanics written in Feynman-Vernon-Hellwarth's torque form \cite{Feynman}, and the formal equivalence of the latter to the time-dependent Schr\"{o}dinger equation for a two-state quantum system \cite{Rangelov09}. Thus the system of equations \eqref{Stokes equation} can be cast into the form of two coupled linear differential equations,
\begin{subequations}\label{2S equations}
\begin{align}
2\text{i}\frac{\text{d}}{\text{d}z}s_{1}(z) &= \Omega _{3}(z)s_{1}(z)+\left[\Omega _{1}(z)-\text{i}\Omega _2(z)\right] s_2(z), \\
2\text{i}\frac{\text{d}}{\text{d}z}s_2(z) &= \left[ \Omega _{1}(z)+\text{i}\Omega _2(z)\right] s_{1}(z)-\Omega _{3}(z)s_2(z),
\end{align}
\end{subequations}
for new complex-valued variables $s_{n}(z)$ $(n=1,2)$. The Stokes vector components are bilinear combinations of these new variables,
\begin{subequations}\label{S-s}
\begin{align}
S_1(z) &=  s_{1}^{\ast }(z)s_2(z)+s_{1}(z)s_2^{\ast }(z), \\
S_2(z) &=  -i\left[ s_{1}^{\ast }(z)s_2(z)-s_{1}(z)s_2^{\ast }(z)\right], \\
S_3(z) &=  \left\vert s_1(z)\right\vert ^2-\left\vert s_{2}(z)\right\vert ^2.
\end{align}
\end{subequations}
Equations \eqref{2S equations} can be written as a single equation,
\begin{equation}\label{2S equationsH}
2\text{i}\frac{\text{d}}{\text{d}z}\mathbf{s}(z)=\mathbf{H}(z)\mathbf{s}(z),
\end{equation}
for the vector $\mathbf{s}(z)=[s_{1}(z),s_2(z)]^{T}$.
Here the Hermitean matrix
\begin{equation}\label{H}
\mathbf{H}(z) = \Omega _{1}(z)\bm{\sigma }_{1}(z)+\Omega _2(z)\bm{\sigma }_2(z)+\Omega _{3}(z)\bm{\sigma }_{3}(z),
\end{equation}
corresponds to the Hamiltonian in a quantum two-state system with $\bm{\sigma}_n$ being the Pauli spin matrices.

\subsection{Solution of the Stokes polarization equation in terms of a two-state solution}

Because there are a number of analytically exactly soluble two-state models, we can use the formal correspondence between the Stokes polarization equation \eqref{Stokes equation} and the two-state equations \eqref{2S equations} and derive exact analytic solutions for the polarization evolution in a birefringent medium. Because the ``Hamiltonian'' \eqref{H} is Hermitean, the evolution matrix (the propagator) for the variables $s_1(z)$ and $s_2(z)$ is unitary, which is conveniently parameterized by four real parameters $a,b,c,d$ as
\begin{equation}
\left[\begin{array}{c} {s}_{1}(z_{f}) \\ {s}_{2}(z_{f})\end{array}\right]
 = \left[ \begin{array}{cc} a+\i b & c+\i d \\ -c+\i d & a-\i b \end{array}\right]
\left[ \begin{array}{c} {s}_{1}(z_{i}) \\ {s}_{2}(z_{i})\end{array}\right] ,
 \label{EMatrixC}
\end{equation}%
 with $a^2+b^2+c^2+d^2=1$. The ``transition probability'' in the two-state problem is $p=c^2+d^2$, whereas $a^2+b^2=1-p$ gives the ``survival probability''. We express the evolution of the Stokes polarization vector in terms of the parameters of the two-state propagator,
\begin{equation}
\left[\begin{array}{c} S_1(z_{f}) \\ S_2(z_{f}) \\ S_3(z_{f}) \end{array}\right]
= \mathbf{U}(z_{f},z_{i})
\left[ \begin{array}{c} S_1(z_{i}) \\ S_2(z_{i}) \\ S_3(z_{i}) \end{array} \right] ,
\end{equation}
where $\mathbf{U} (z_{f},z_{i})$ is found by using Eqs.~ \eqref{S-s} and \eqref{EMatrixC},
\begin{align}\label{Stokes propagator}
&\mathbf{U} (z_{f},z_{i})=\notag\\
&\left[ \!\! \begin{array}{ccc}
a^2-b^2-c^2+d^2 & -2(ab+cd) & 2(ac-bd) \\
2(ab-cd) & a^2-b^2+c^2-d^2 & 2(ad+bc) \\
-2(ac+bd) & -2(ad-bc) & a^2+b^2-c^2-d^2
\end{array}\!\!\right].
\end{align}


Equation \eqref{Stokes propagator} allows us to immediately write down the conditions for several representative important transformations of light polarization.

(i) Transformation from right circular, $\mathbf{S}(z_i)=[0,0,1]^T$, to left circular polarization, $\mathbf{S}(z_f)=[0,0,-1]^T$ (or vice versa), requires:
\begin{equation}
a =b =0
\end{equation}
which corresponds to a transition probability $p=1$ in the two-state quantum system.

(ii) Transformation from right circular, $\mathbf{S}(z_i)=[0,0,1]^T$, to linear polarization (or vice versa), $\mathbf{S}(z_f)=[\cos 2\phi,\sin 2\phi,0]^T$, requires:
 \begin{equation}
a^2+b^2=c^2+d^2=\frac12
\end{equation}
which corresponds to a transition probability $p=\frac12$ in the two-state quantum system.

\section{Models with exact analytic solutions} \label{Sec3}

Equation \eqref{Stokes propagator} allows us to use the available exact analytic solutions for two-state quantum systems
 to derive exact analytic solutions for the evolution of the polarization of light propagating through a birefringent medium.
We select here three such models: the one-dimensional (resonance) model, the popular Landau-Zener model \cite{Landau,Zener} (in its finite version), and the flexible Demkov-Kunike model \cite{DK,VitanovARZ} (also in its finite version).

\subsection{One-dimensional model}

We assume that the birefringent medium stretches from $z_{i}$ to $z_{f}$. The elements of the birefringence vector for this model read
\begin{equation}\label{Res model}
\Omega _1(z) = f(z/L) , \quad
\Omega _2(z) = \Omega _3(z) = 0,
\end{equation}
where $f(z/L)$ is any function of $z$ for $z_{i} \le z \le z_{f}$ and $L$ is a length scale parameter. 
For example, $f(z/L)$ can be chosen in such a way that $a=b=0$, so we can have transformation from left circular to right circular polarization, which is known as a half-wave plate \cite{Wolf}.

\begin{figure}[tb]
 \includegraphics[width=0.45 \textwidth]{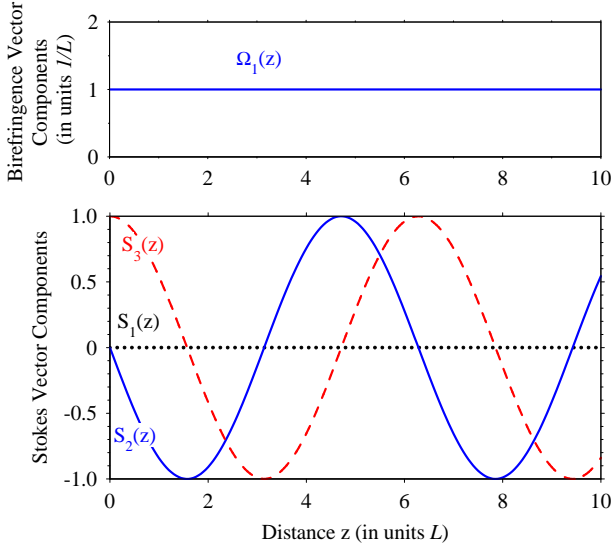}
 \caption{Transformation of right circular polarization \eqref{right} with a birefringence vector $\mathbf{\Omega}(z)=[f(z), 0, 0]^{T}$ for the one-dimensional model, where $f(z/L)=1/L$. The polarization evolution corresponds to Rabi flopping in atomic physics (Eq. \eqref{Res_prop}).}
 \label{ResGraph}
\end{figure}

This one-dimensional model corresponds to exact resonance in a two-state system, which causes Rabi oscillations \cite{Sho90}, and for which the propagator parameters are well known,
\begin{subequations}\label{Res parameters}
\begin{align}
a &=\cos(A/2) , \\
d &= -\sin(A/2) , \\
b &= c = 0,
\end{align}
\end{subequations}
where $A (z_{f},z_{i})= \frac{1}{L}\int_{z_{i}}^{z_{f}} \! f(z /L) \, \mathrm{d}z$.
Hence the evolution matrix for the Stokes vector is
\begin{equation} \label{Res_prop}
\mathbf{U} (z_{f}, z_{i}) =
\left[ \begin{array}{ccc}
1 & 0 & 0 \\
0 & \cos(A) & -\sin(A) \\
0 & \sin(A) & \cos(A)
\end{array}\right].
\end{equation}
The transformation from right circular to left circular polarization is achieved when $A=(2k+1) \pi$, $k=0,1,2..$, i. e. then $a=b=0$. An example with the simplest function $f(z/L)=1/L$, i. e. a constant, is shown in  Fig. \ref{ResGraph}. In this case,  $A=(z_f -z_i) /L$ and the transformation is achieved for $z_f -z_i= (2k+1) \pi L$.

\subsection{Finite Landau-Zener model}

In this exactly soluble model the birefringence vector components are given by
\begin{equation}\label{LZ model}
\Omega _{1}(z) = \Omega_{0}, \quad \Omega _2(z) =0, \quad \Omega _{3}(z) = \beta^{2} z,
\end{equation}
for $z_{i} \le z \le z_{f}$, and $\Omega_{k} = 0$ ($k=1,2,3$) otherwise. These elements correspond to the finite Landau-Zener model in atomic physics for the dynamics of the new variables $s_{1}(z)$ and $s_{2}(z)$, as defined in Eqs.~\eqref{S-s} \cite{Zener,VitanovLZ}.
For this model, we find the propagator parameters in a similar fashion as in \cite{VitanovLZ},
\begin{subequations}\label{LZ propagator}
\begin{align}
a+\i b = & \frac{\Gamma(1-i\alpha^{2})}{\sqrt{2\pi}} [ D_{i\alpha^{2}} (\zeta_{f})D_{-1+i\alpha^{2}}(-\zeta_{i}) \notag\\
 & + D_{i\alpha^{2}}(-\zeta_{f}) D_{-1+i\alpha^{2}}(\zeta_{i}) ] \\
c+\i d = & \frac{\Gamma(1-i\alpha^2)}{\alpha \sqrt{2\pi}} e^{i \frac{\pi}{4}} [ D_{i\alpha^2} (-\zeta_{f})D_{i\alpha^{2}}(\zeta_{i}) \notag\\
 & - D_{i\alpha^{2}}(\zeta_{f}) D_{i\alpha^{2}}(-\zeta_{i}) ].
\end{align}
\end{subequations}
where $D_{\nu} (z)$ is the parabolic cylinder (Weber) function \cite{VitanovLZ,AS} and $\zeta(z) = \beta z\e^{-i\frac{\pi}{4}} $, i. e. $\zeta_{i,f} = \beta z_{i,f} e^{-i\frac{\pi}{4}}$; $\alpha = {\Omega_0}/{2\beta}$. We find the propagator for the evolution of the Stokes polarization vector from Eq. \eqref{LZ propagator} by using Eq. \eqref{Stokes propagator}.

\begin{figure}[tb]
\includegraphics[width=0.45 \textwidth]{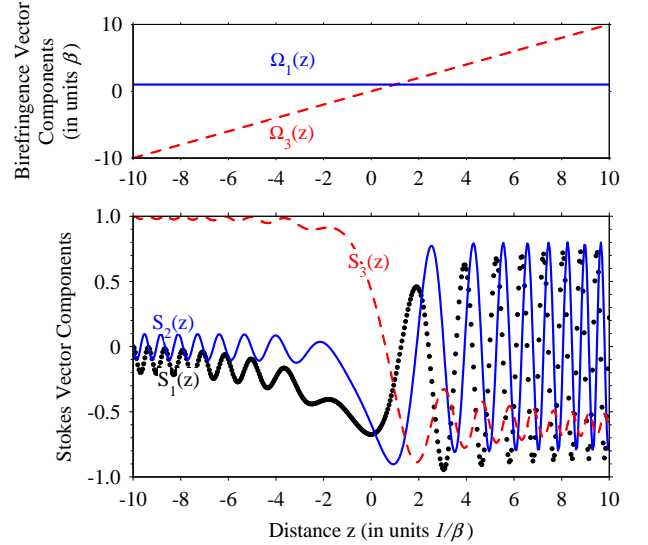}
\caption{Transformation of right circular polarization \eqref{right} with a birefringence vector for the Landau-Zener model with $\Omega_{0}=\beta$ and $z_i=-10 /\beta$ (Eq. \eqref{LZ propagator}, Eq. \eqref{LZ_null} can also be applied for $z \rightarrow +\infty $).}
\label{LZRL}
\end{figure}
In the limiting case when $L=z_f=-z_i\gg 1/\beta$ (``symmetric crossing'') we use the ``weak coupling'' asymptotic of $D_\nu(\zeta)$ (for fixed $\nu$ and $|\zeta|\to\infty$) \cite{VitanovLZ}:
\begin{subequations}\label{LZSym}
\begin{align}
D_\nu(\zeta) & \sim \zeta^\nu e^{-\zeta^2/4} [1+\mathcal{O}(|\zeta|^{-2})], \quad (|\arg \zeta|<3\pi/4), \\
D_\nu(\zeta) & \sim  e^{i\pi\nu} (-\zeta)^\nu e^{-\zeta^2/4}  [1+\mathcal{O}(|\zeta|^{-2})] \notag \\
 &+ \frac{\sqrt{2\pi}}{\Gamma(-\nu)} e^{i\pi(\nu+1)/2} (-i\zeta)^{-1-\nu} e^{z^2/4}  [1+\mathcal{O}(|\zeta|^{-2})], \notag \\
 &\quad (|\arg \zeta|\geqq 3\pi/4),
\end{align}
\end{subequations}
and we find the following approximation of the elements of the propagator \cite{VitanovLZ}:
\begin{subequations}
\begin{align}
a+\i b &= e^{-\pi \alpha^{2}}, \\
c+\i d &= -\sqrt{1-e^{-2\pi \alpha^{2}}}\,  e^{\i\phi}, \\
\phi &= \frac {\pi}{4} + \frac{\beta^2 L^2}{2} + {\alpha^2} \ln (\beta^2 L^2) + \arg \Gamma(1- i\alpha^{2}).
\end{align}
\end{subequations}
This case is illustrated in Fig. \ref{LZRL}, where the exact solution based on \eqref{LZ propagator} approaches the asymptotic one \eqref{LZSym} for $L=z_f=-z_i\gg 1/\beta$.

\begin{figure}[tb]
 \includegraphics[width=0.45 \textwidth]{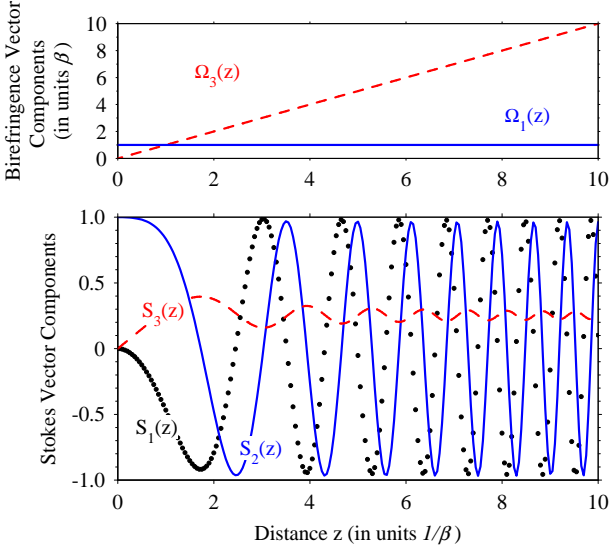}
\caption{Transformation of linear polarization \eqref{linear}, $\phi =\pi /4$,  with a birefringence vector for the Landau-Zener model with $\Omega_{0}=\beta$ and $z_i=0$ (Eq. \eqref{LZ propagator}, Eq. \eqref{LZ_null} can also be applied for $z \rightarrow +\infty $).}
 \label{LZRL2}
\end{figure}
Another important limiting case is when $z_i=0$ and $z_f=L\gg 1/\beta$ (``half crossing''). Then, we use $D_\nu(0)=2^{\nu/2}\sqrt{\pi}/ \Gamma(\frac{1- \nu}{2})$ to simplify Eq. \eqref{LZ propagator} \cite{VitanovLZ}
\begin{subequations}\label{LZ_null}
\begin{align}
a+\i b = & \frac{\Gamma(1-i\alpha^{2})}{2 \Gamma(1- \frac{i\alpha^{2}}{2})} 2^{i\alpha^2/2} [ D_{i\alpha^{2}} (\zeta_{f})  + D_{i\alpha^{2}}(-\zeta_{f}) ] \\
 \sim & \frac{\Gamma(1-i\alpha^{2})}{ \Gamma(1- \frac{i\alpha^{2}}{2})}  e^{-\pi\alpha^2/4 + i\chi} \cosh(\pi\alpha^2/2),  \\
c+\i d = & \frac{\Gamma(1-i\alpha^2)}{\alpha \sqrt{2} \Gamma(\frac{1- i\alpha^{2}}{2})} e^{i \frac{\pi}{4}} 2^{i\alpha^2/2}  [ D_{i\alpha^2} (-\zeta_{f})  - D_{i\alpha^{2}}(\zeta_{f}) ],\\
 \sim & \frac{\sqrt{2}\, \Gamma(1-i\alpha^2)}{\alpha \Gamma(\frac{1- i\alpha^{2}}{2})} e^{-\pi\alpha^2/4 + i\chi -3 i \pi/4} \sinh(\pi\alpha^2/2).
\end{align}
\end{subequations}
with $\chi= \beta^2 L^2/4 + \alpha^2\ln(\beta L\sqrt{2})]$. This case is illustrated in Fig. \ref{LZRL2}, where the exact solution based on \eqref{LZ propagator} approaches the asymptotic one \eqref{LZSym} for $z_i=0$ and $z_f=L\gg 1/\beta$.

\subsection{Finite Demkov-Kunike model}

We shall derive an analytic solution for birefringence vector components given by:
\begin{subequations}
\label{model}
\begin{align}
\Omega _{1}(z)& =\Omega _{0} \text{sech}(z/L), \\
\Omega _2(z)& =0, \\
\Omega _{3}(z)& =\Delta _{0}  +B _{0} \tanh (z/L),
\end{align}%
\end{subequations}
for $z_{i} \le z \le z_{f}$, and $\Omega_{k} = 0$ $(k=1,2,3)$, otherwise. These elements correspond to the finite Demkov-Kunike model for the dynamics of the variables $s_{1}(z)$ and $s_{2}(z)$, as defined in Eq. \eqref{S-s} \cite{DK}. The solution is derived in a similar fashion to the infinite Demkov-Kunike model in atomic physics \cite{DK}, with the additional assumption of finite initial and final conditions.

\begin{figure}[tb]
 \begin{center}
  \includegraphics[width=0.45  \textwidth]{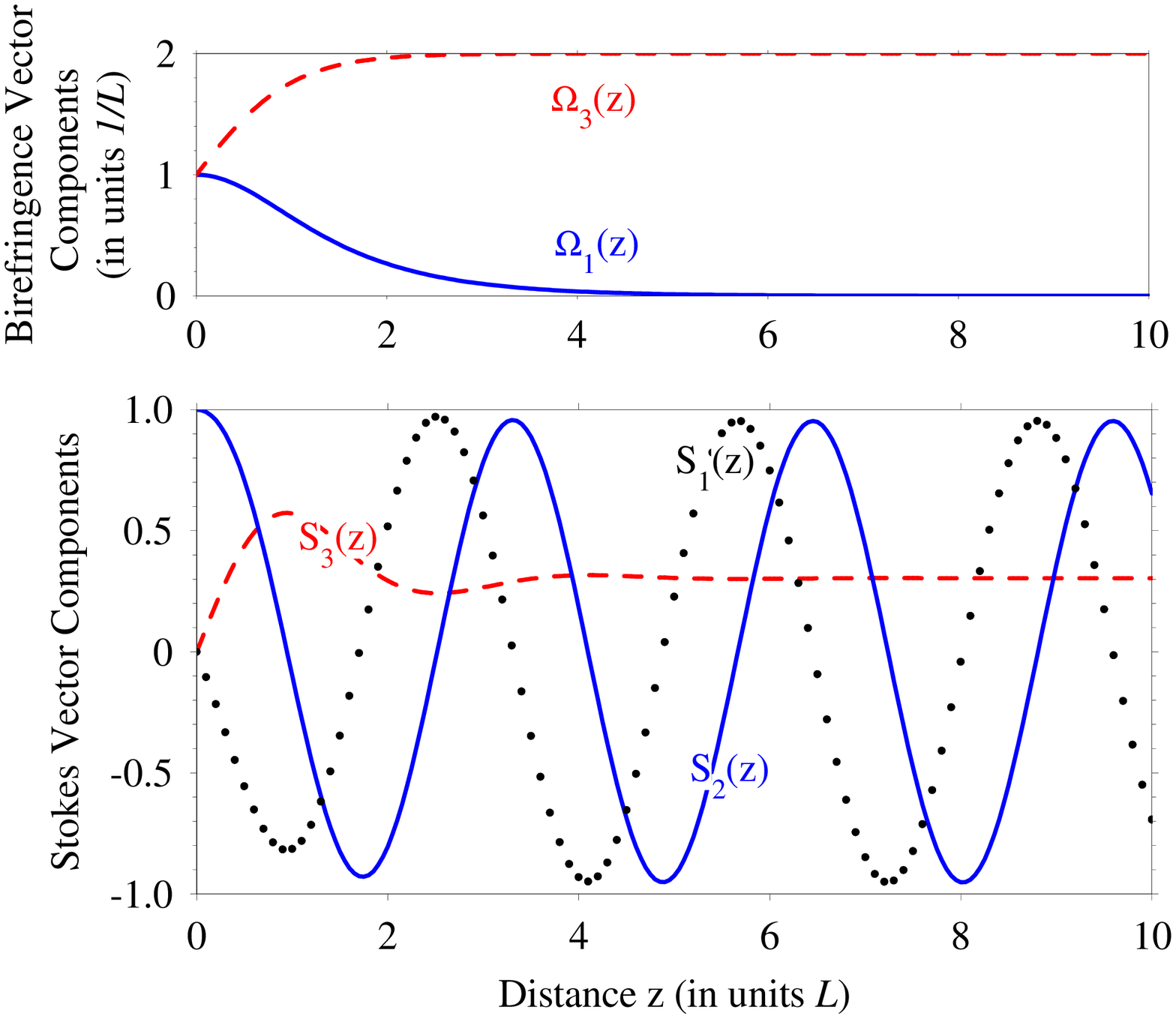}
 \end{center}
 \vspace{-.3cm}\caption{Transformation of linear polarization \eqref{linear}, $\phi =\pi /4$, with a birefringence vector for the Demkov-Kunike model with $\Omega _{0}=1/L$, $\Delta _{0}=1/L$, $B _{0}=1/L$, and $z_i=-10 L$  (Eq. \eqref{DK prop1}, Eq. \eqref{DK prop3} can also be applied for $z \rightarrow +\infty $).}
 \label{DKLin}
\end{figure}

The propagator parameters for this model for any $z_{i}$ and $z_{f}$ are:
\begin{subequations}\label{DK prop1}
\begin{align}
a+\i b = &e^{i\frac{\Phi(z_{f}, z_{i})}{2}} [F_{1}(z_{f})F_{1}^{*}(z_{i})+\frac{|\alpha|^{2} 2^{-4 \Im{(\beta)}}}{|1 -\nu|^{2}} \notag \\
& \times F_{2}(z_{f}) F_{2}^{*}(z_{i}) \xi(z_{f})^{1-\nu} \xi(z_{i})^{\nu}]  \\
c+\i d = &\frac{-i \alpha 2^{2i \beta}}{1- \nu} e^{i(\frac{\Phi(z_{f}, z_{i})}{2} - \Phi_{0})} \notag\\
& \times [ F_{2}(z_{f})F_{1}(z_{i}) \xi(z_{f})^{1-\nu} - F_{1}(z_{f})F_{2}(z_{i}) \xi(z_{i})^{1-\nu} ]
\end{align}
\end{subequations}
where $\alpha =\Omega_0L/2$, $\beta = B_0L/2$, $\delta = \Delta_0L/2$, $\lambda = \sqrt{\alpha ^2-\beta ^2}-i\beta$,
$\mu =-\sqrt{\alpha^2-\beta ^2}-i\beta $, $\nu = \frac12+i(\delta -\beta )$,
$\Phi_{0} =\int_{z_{i}}^{0}\Omega_{3} (z^{\prime })dz^{\prime }$, and $\Phi(z,z_{i})=\int_{z_{i}}^{z}\Omega_{3} (z^{\prime })dz^{\prime }$. Furthermore,  $F_{1}(z) \equiv F(\lambda, \mu, \nu, \xi (z))$ and $F_{2}(z) \equiv F(\lambda+1-\nu, \mu+1-\nu, 2-\nu, \xi(z))$ are hypergeometric functions, where $\xi(z) = [\tanh(z/L)+1]/2$ \cite{AS}.

The formulas for the elements of the propagator for any $z_{i}$ are a generalization of the ones derived in \cite{VitanovStenholm} for the special case when $z_{f}=-z_{i}$ and $\beta=0$ (finite Rosen-Zener model). Then, the propagator for the evolution of the Stokes polarization vector is found from here by using Eq. \eqref{Stokes propagator}.

\begin{figure}[tb]
 \includegraphics[width=0.45  \textwidth]{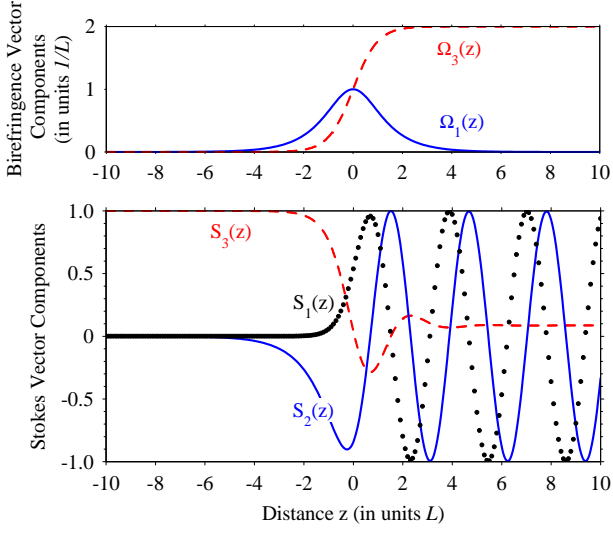}
\caption{Transformation of right circular polarization \eqref{right} with a birefringence vector for the Demkov-Kunike model with $\Omega _{0}=1/L$, $\Delta _{0}=1/L$, $B _{0}=1/L$, and $z_i=0$  (Eq. \eqref{DK prop1}, Eq. \eqref{DK prop2} can also be applied for $z \rightarrow +\infty$).}
 \label{DKRL}
\end{figure}

When $z_{i}=0$ and $z_{f} \rightarrow +\infty$, the elements of the propagator simplify to \cite{BTT}:
\begin{subequations}\label{DK prop2}
\begin{align}
a+\i b &= e^{i\frac{\Phi(z_{f}, 0)}{2}} \tilde{F}_{1} \\
c+\i d &= \frac{-i \alpha}{2 (\nu -\lambda -\mu) } e^{i\frac{\Phi(z_{f}, 0)}{2}} \tilde{F}_{2}
\end{align}
\end{subequations}
where $\tilde{F}_{1} \equiv F(-\lambda, -\mu, \nu -\lambda -\mu, \frac{1}{2})$, $\tilde{F}_{2} \equiv F(1 -\lambda, 1 -\mu, 1 + \nu -\lambda -\mu, \frac{1}{2})$. An example for the manipulation of polarization by a birefringence vector with the characteristics of the Demkov-Kunike model and $z_{i}=0$ is given Fig. \ref{DKLin}.

The formulas for the elements of the propagator can be presented with elementary functions in the case of the original Demkov-Kunike model when $z_i \rightarrow -\infty$ and $z_f \rightarrow +\infty$. Then, the general formula for the elements of the propagator can be expressed by \cite{DK}:
\begin{subequations}\label{DK prop3}
\begin{align}
a+\i b =&  e^{i(\delta(z_f - z_i)/L  + \beta(z_f + z_i)/L)} \\
& \times \frac{\Gamma(\nu) \Gamma (\nu -\lambda-\mu)}{\Gamma(\nu-\lambda) \Gamma (\nu-\mu)}\\
c+\i d =& \frac{-i \alpha 2^{2i \beta}}{1- \nu} e^{i(\delta (z_f + z_i)/L + \beta(z_f - z_i)/L -2\beta\ln{2})}\\
& \times \frac{\Gamma(2-\nu) \Gamma (\nu -\lambda-\mu)}{\Gamma(1-\lambda) \Gamma (1-\mu)}
\end{align}
\end{subequations}
An example for such manipulation of polarization is shown in Fig. \ref{DKRL}.

\subsection{Adiabatic limit}

\begin{figure}[tb]
 \includegraphics[width=0.45 \textwidth]{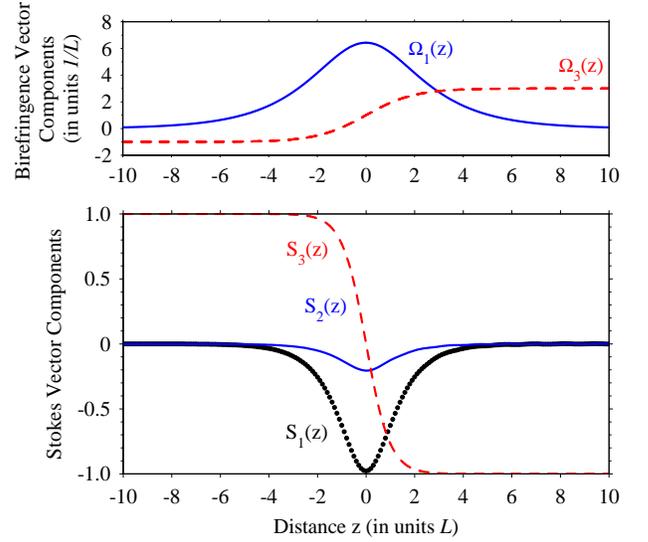}
 \caption{Transformation of right circular polarization \eqref{right} to left circular polarization \eqref{left} with a birefringence vector for the Demkov-Kunike model with parameters, which satisfy the adiabatic condition \eqref{AdCond}: $\Omega _{0}=5/L$, $\Delta _{0}=0$, $B _{0}=5/L$ (Eq. \eqref{DK prop1}).}
 \label{DK_RAP}
\end{figure}

In the limit of large rotary power (which corresponds to large pulse area in quantum physics),
\begin{subequations} \label{AdCond}
\begin{align}
\text{Global condition:} & \int_{z_{i}}^{z_{f}}|\mathbf{\Omega }(z)| \mathrm{d}z \gg 1, \\
\text{Landau-Zener model:} & ~\Omega _{0} \gg ~\beta \\
\text{Demkov-Kunike model:} & ~\Omega _{0} ~\ge B_{0} \ge \frac{1}{L} .
\end{align}
\end{subequations}
the evolution of the Stokes polarization vector becomes adiabatic. Additionally, both the Landau-Zener and Demkov-Kunike models also involve a ``level crossing'' and they are therefore examples of exactly solvable models of rapid adiabatic passage (RAP) \cite{AE,Sho08,Vit01b,Vit01a}.

\begin{figure}[tb]
 \hspace{-0.3cm}\includegraphics[width=0.4 \textwidth]{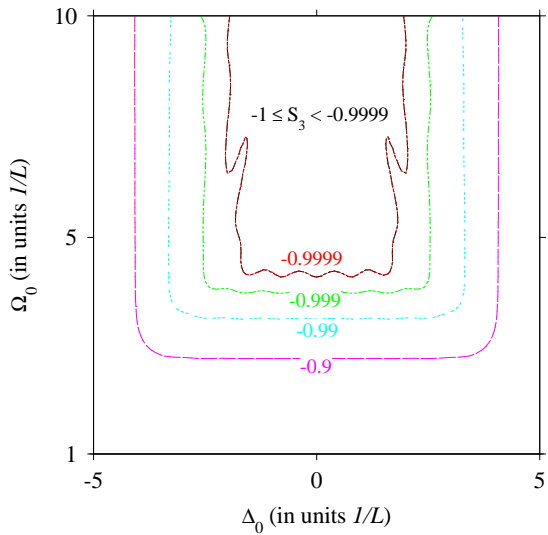}
 \caption{Contourplot of $S_{3}(z_{f}=10 L)$ where the birefringence vector corresponds to the Demkov-Kunike model dynamics of the corresponding two-state system with $\mathbf{S}(z_{i}=-10 L)=[0, 0, 1]^{T}$ with parameters: $B _{0}=5 /L$. Left circular polarization is achieved when $S_{3}(z_{f})=-1$ (Eq. \eqref{DK prop1}).}
 \label{C_RAP}
\end{figure}

RAP traditionally provides a well-studied robust and efficient
method for producing complete population transfer between two
bound states of a quantum system
\cite{AE,Sho08,Sho90,Vit01b,Vit01a}. The population change is
achieved by sweeping the carrier frequency of a laser pulse
through resonance with a two-state system (parameterized by
detuning), while simultaneously pulsing the strength of the
interaction (parameterized as time-varying Rabi frequency). The
sweep of detuning corresponds to a crossing of the so-called
diabatic energy curves. When the interaction changes sufficiently
slowly, so that the time evolution is adiabatic, there will occur
a complete population transfer from the ground state to the
excited state.

In terms of manipulation of polarization, RAP is a broadband technique for robust adiabatic conversion of light polarization and it is therefore applicable to a wide range of frequencies.
It is also robust to variations in the propagation length, the rotary power and other parameters variations. The robustness of the RAP polarization conversion with respect to deviation in the propagation length for the Demkov-Kunike model is demonstrated in Fig. \ref{DK_RAP}. This figure shows that the Stokes vector is quite stable with regard to changes in $z$ for $z \rightarrow +\infty$ in contrast to the previous figures where it oscillates with $z$. The robustness to variations in the parameters of the birefringence vector are demonstrated in Fig. \ref{C_RAP}, which is a contourplot of $S_{3}(z_{f})$ with respect to the parameters $\Delta_{0}$ and $\Omega _{0}$, while the other paremeters are fixed. As we can see, right circular polarization is transformed to a left circular one for quite large deviations of the two parameters.

\section{Conclusion} \label{Sec4}

In this paper we have used the analogy between the torque equation of motion and the coherent atomic excitation of a two state system to describe the evolution of the Stokes polarization vector in a birefringent medium without polarization dependent loss. We have given exact analytic solutions for this evolution when the birefringent vector corresponds to the dynamics of quantum two-state systems, which are described by the one-dimensional (resonance) model, the Landau-Zener model, and the Demkov-Kunike model. Finally, we have demonstrated how this approach could be applied for broadband conversion of light polarization in analogy to rapid adiabatic passage (RAP) in two-state quantum systems.

This approach can be useful in many other branches of physics where we need exact analytic solutions to a torque equation. These include the description of the Newton equation of motion, the classical motion of a charged particle in oscillating magnetic and electric fields, e. g. for charged particle confinement, optical waveguides, polychromatic beam splitter, sum frequency conversion techniques in nonlinear optics, the dynamics of a spin in a magnetic field, the behaviour of a gyroscope acted on by gravity, etc.

\acknowledgments
This work is supported by the European Commission network FASTQUAST and the Bulgarian NSF grants VU-I-301/07, D002-90/08 and DMU02-19/09.


\end{document}